\documentclass[reprint,NumberedRefs]{JASA}
\usepackage{enumitem}
\usepackage{ragged2e}
\usepackage{setspace}
\usepackage{cleveref}
\usepackage{multirow}
\usepackage{upgreek}
\def\equationautorefname~#1\null{(#1)\null}

\newcommand{\AlgBlankLine}{\par\medskip}
\usepackage[all]{hypcap}

\begin{document}

\title{Computationally efficient full-waveform inversion of the brain using frequency-adaptive grids and lossy compression}
\author{Letizia Protopapa}
\email{letizia.protopapa19@imperial.ac.uk}
\affiliation{Department of Bioengineering, Imperial College London, London, SW7 2AZ, United Kingdom}
\author{Carlos Cueto}
\email{c.cueto@imperial.ac.uk}
\affiliation{Department of Bioengineering, Imperial College London, London, SW7 2AZ, United Kingdom}

\date{\today}

\begin{abstract}
\vspace{3mm}
A tomographic technique called full-waveform inversion has recently shown promise as a fast, affordable, and safe modality to image the brain using ultrasound. However, its high computational cost and memory footprint currently limit its clinical applicability. Here, we address these challenges through a frequency-adaptive discretisation of the imaging domain and lossy compression techniques. Because full-waveform inversion relies on the adjoint-state method, every iteration involves solving the wave equation over a discretised spatiotemporal grid and storing the numerical solution to calculate gradient updates. The computational cost depends on the grid size, which is controlled by the maximum frequency being modelled. Since the propagated frequency typically varies during the reconstruction, we reduce reconstruction time and memory use by allowing the grid size to change throughout the inversion. Moreover, we combine this approach with multiple lossy compression techniques that exploit the sparsity of the wavefield to further reduce its memory footprint. We explore applying these techniques in the spatial, wavelet, and wave atom domains. Numerical experiments using a human-head model show that our methods lead to a 30\% reduction in reconstruction time and up to three orders of magnitude less memory, while negligibly affecting the accuracy of the reconstructions.  
\end{abstract}

\maketitle


\section{\label{sec:intro} Introduction}

Neurological disorders currently represent a major cause of death and disability worldwide, with stroke being one of the largest contributors \citep{neuro1}. In the case of stroke, appropriate treatment can only be provided after the brain has been imaged, with treatment delays having a significant impact on patient outcomes \citep{neuro2}. Therefore, a portable and fast neuroimaging technique that allows diagnosis and treatment to be performed at the point of first contact has the potential for significant impact in clinical neurological practice. Recently, it has been shown that this could be achieved through ultrasound tomography with full-waveform inversion (FWI), a technique originally developed to image the Earth’s subsurface\citep{lluis}. While conventional ultrasound is unable to image the brain across the adult human skull, FWI can do this with a high spatial resolution by relying on an accurate description of the physics of wave propagation\citep{lluis}. However, FWI currently exhibits high computational costs, both in terms of time and memory requirements, with realistic 3D imaging problems taking tens of hours to compute in high-performance computer clusters and requiring hundreds of gigabytes in memory\citep{lluis, bachmann}. Given that imaging speed is crucial for prompt diagnosis of diseases like stroke, increased computational efficiency is fundamental to enable the translation of brain FWI to clinical practice.

In geophysics, previous studies have attempted to overcome these computational limitations through different approaches. Some of these studies have focused on the fact that, in FWI, most of the computational time is spent modelling wave propagation \citep{vargrid, hursky}, which requires the full numerical solution of the acoustic wave equation over discrete grids. The size of the grid is controlled by the acoustic properties of the imaging target and the temporal frequencies involved. More specifically, the grid spacing is proportional to the minimum velocity in the model and inversely proportional to the maximum frequency being modelled. Even though both frequencies and velocities vary during the inversion, a constant grid spacing is typically used for the whole process, which is determined by the highest inverted frequency and minimum imaged velocity. Because the grid spacing could be larger during the inversion of low frequencies, using a fixed grid results in oversampling and redundant computations. For this reason, previous studies have explored the use of adaptive grids, where the grid spacing changes with the background acoustic velocity \citep{vargrid, multiscale} or the frequency of the propagated wave \citep{kormann}, thus allowing the use of a lower number of grid points with respect to conventional FWI. The use of adaptive grids requires repeatedly resampling the model being reconstructed, for which it is common to rely on standard resampling methods such as linear and spline interpolation. However, traditional resampling algorithms perform poorly in the presence of high-contrast areas such as the head, in which the speed of sound and density of the skull are significantly higher than those of the surrounding tissues.

While the size of the grid also has an impact on memory consumption, the large memory requirements of FWI are due to the use of the adjoint-state method, which allows us to efficiently calculate the gradient of the misfit function\citep{hursky}. The adjoint-state method requires solving the acoustic wave equation forward in time and then storing this solution in memory at all time steps until the adjoint wave equation is solved because the two wavefields must be accessed simultaneously for computing the gradient. Prior studies in the field of geophysics have explored ways to reduce FWI memory consumption through different forms of checkpointing\citep{kukreja1, timerev} and lossy compression\citep{kukreja1, boehm}. In checkpointing approaches, the forward solution of the wave equation is stored only at specific time steps, while the remaining ones are discarded and later recomputed during the adjoint calculation\citep{kukreja1}. This, however, can result in significant computational overhead. For this reason, checkpointing is sometimes combined with other techniques, such as lossy compression, to reduce the size of the stored states and, therefore, the amount of states to be recomputed\citep{kukreja1}. Previous studies have also used lossy compression on its own to reduce the size of the stored wavefield, thus avoiding the large computational overhead that characterises checkpointing methods\citep{boehm}. 

Here, we propose a combination of techniques aimed at reducing the computational cost of FWI while preserving imaging accuracy in the presence of high-contrast regions such as the human head. More specifically, we present a multi-grid approach, together with an edge-preserving interpolation algorithm, that allows the grid size to vary with the inverted frequencies, thus reducing both reconstruction time and memory use. Moreover, we take advantage of the inherent sparsity of the wavefield to further reduce its memory footprint through multiple lossy compression techniques, including hard thresholding and a sparse, half-precision representation of the wavefield. We compare the effect of assuming sparsity in three different domains: the spatial, wavelet, and wave atom\citep{wa} domains. We present results of applying our methodology to image a numerical phantom of the human head, showing that it leads to a 30\% shorter reconstruction time and up to three orders of magnitude less memory.

The rest of the paper is structured as follows: first, we introduce our proposed algorithms for multi-grid FWI and lossy compression in detail, as well as the numerical tests we carried out using a phantom of the human head. Then, we present the results of our experiments, showing how our methodology significantly reduces both computational time and memory use. Finally, we provide a discussion of our work and present our conclusions.

\section{\label{sec:Methods} Methods}
\subsection{Full-waveform inversion}

As previously introduced, FWI is a technique that seeks to reconstruct the acoustic properties (generally speed of sound) of an object by solving an associated physics-constrained inverse problem. This involves starting from a model that is close enough to the object of the reconstruction to ensure convergence, and updating it iteratively by minimising the misﬁt between a set of experimentally observed and numerically predicted data\citep{operto}. The gradient of this cost function is obtained through the adjoint-state method, which involves solving the wave equation forward in time, storing the solution in memory, and then solving the adjoint wave equation backwards in time\citep{operto}.

Solutions of the wave equation are obtained over a discrete spatiotemporal grid by using numerical methods such as finite differences (FD), as in this study, or finite elements. The spatial and temporal steps of the discretisation are governed by dispersion and stability limits respectively. In particular, the maximum spatial grid spacing is given by,
\begin{eqnarray} 
dx_{max}=\frac {\lambda_{min}}{n}=\frac {v_{min}}{n\,f_{max}}
\label{eq:dx_max}
\end{eqnarray}
\noindent
where $\lambda_{min}$ is the minimum wavelength within the model, i.e. the ratio of the minimum wave velocity $v_{min}$ and the maximum frequency $f_{max}$, while $n$ is a positive number indicating the grid points per wavelength, which depends on the implemented numerical scheme. In our implementation of the wave equation, we consider $n=5$. Similarly, the maximum temporal spacing of our discretisation is determined by the Courant-Friedrichs-Lewy (CFL) condition\citep{cfl},
\begin{eqnarray} 
dt_{max}=\frac {\mu\,dx_{max}}{v_{max}}
\label{eq:cfl}
\end{eqnarray}
\noindent
where $v_{max}$ is the maximum velocity in the model and $\mu$ is a factor that depends on the dimensionality of the wave equation and the accuracy of the numerical approximations. In this study, we set $\mu$ to $0.55$\citep{sampling}.

Due to the oscillatory nature of the data, the solution of the inverse problem can lead to a local minimum rather than the global minimum, a phenomenon known as cycle-skipping. To reduce the risk of non-convergence, a multiscale approach is commonly used\citep{bunks}, in which the inversion is performed in stages, starting from the low frequency components of the data (which are less sensitive to cycle-skipping) and gradually introducing the higher frequencies into the optimisation process.

\subsection{\label{subsec:multi-grid} Multi-grid approach}

\begin{figure}[ht]
\includegraphics[width=\reprintcolumnwidth]{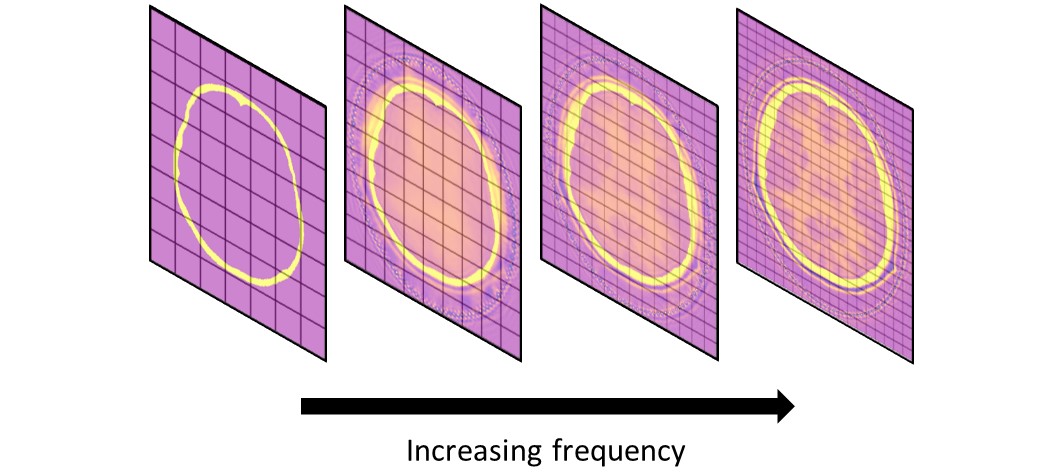}
\vspace{-5mm}
\caption{\label{fig:FIG1}{The spatial grid becomes finer as higher frequencies are introduced into the inversion process.}}
\vspace{-3mm}
\end{figure}

Because frequencies are introduced gradually into the FWI reconstruction, and the grid spacing is controlled by the maximum frequency being propagated, we can reduce the computational effort of solving each FWI iteration by adopting a different grid for each frequency band (Fig.~\ref{fig:FIG1}). This also results in lower memory requirements, due to the smaller size of the wavefields being stored at low frequencies. 

To obtain a different grid for each band, the model being reconstructed must be resampled whenever a new frequency band is introduced. To correctly resample the model, we need to account for the large acoustic contrast that can arise in certain regions of the model, such as between the skull and the surrounding soft tissue. In these high-contrast regions, using a high-order interpolation without considerable prior smoothing results in ringing artefacts\citep{feng}, while a low-order interpolation causes excessive blurring. To avoid both these problems, we have developed an edge-preserving interpolation algorithm. 

\begin{figure}[b]
\vspace{1mm}
\includegraphics[width=\reprintcolumnwidth]{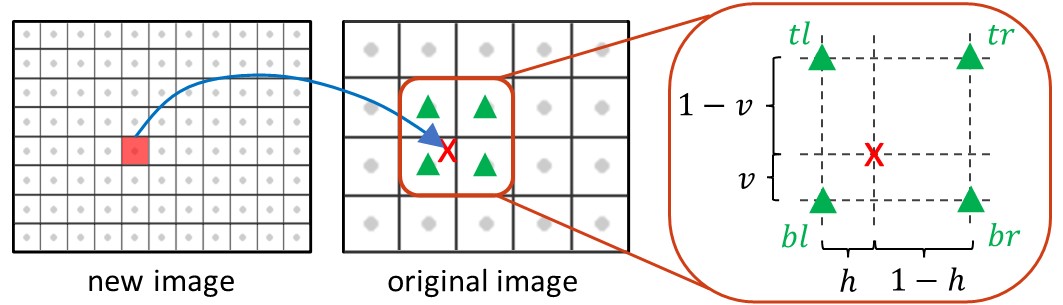}
\vspace{-5mm}
\caption{\label{fig:FIG2}{An example of interpolation showing how a grid point in the new image corresponds to a position that might be off-grid in the original image. Such position is marked by a cross, while the four grid points surrounding it are shown as triangles. The close-up on the right displays the names given to the intensities of the grid points as well as to their distances in the horizontal and vertical direction with respect to the interpolated point.}}
\end{figure}

The algorithm starts by determining which grid points of the original image are close to or belong to an edge using the Sobel edge detector. Then, for each grid point in the new image, it determines the (possibly off-grid) coordinates that indicate the position of this grid point in the original image (marked by a cross in Fig.~\ref{fig:FIG2}). After that, the algorithm considers the four grid points that surround such position in the original image (shown as triangles in Fig.~\ref{fig:FIG2}) and calculates the value $X_b$ predicted by bilinear interpolation (Algorithm  \hyperlink{algo:bilinear}{1}) at that point.

\vspace{1mm}
\begin{algorithm}[h]
\hypertarget{algo:bilinear}{}
\caption{BilinearInterpolation}
\label{alg:BilinearInterpolation}
\begin{algorithmic}[1]
\footnotesize
\REQUIRE{The intensities of the grid points ($br$, $bl$, $tr$, $tl$) and the distances $h$ and $v$, which appear in Fig.~\ref{fig:FIG2}}
\AlgBlankLine
\ENSURE{The interpolated intensity}
\AlgBlankLine
\STATE $b \gets h*br+(1-h)*bl$
\STATE $t \gets h*tr+(1-h)*tl$
\STATE $output \gets v*t+(1-v)*b$
\AlgBlankLine
\end{algorithmic}
\end{algorithm}
\vspace{6mm}

\begin{figure}[h]
\vspace{2mm}
\hypertarget{fig:FIG3}{}
\includegraphics[width=\reprintcolumnwidth]{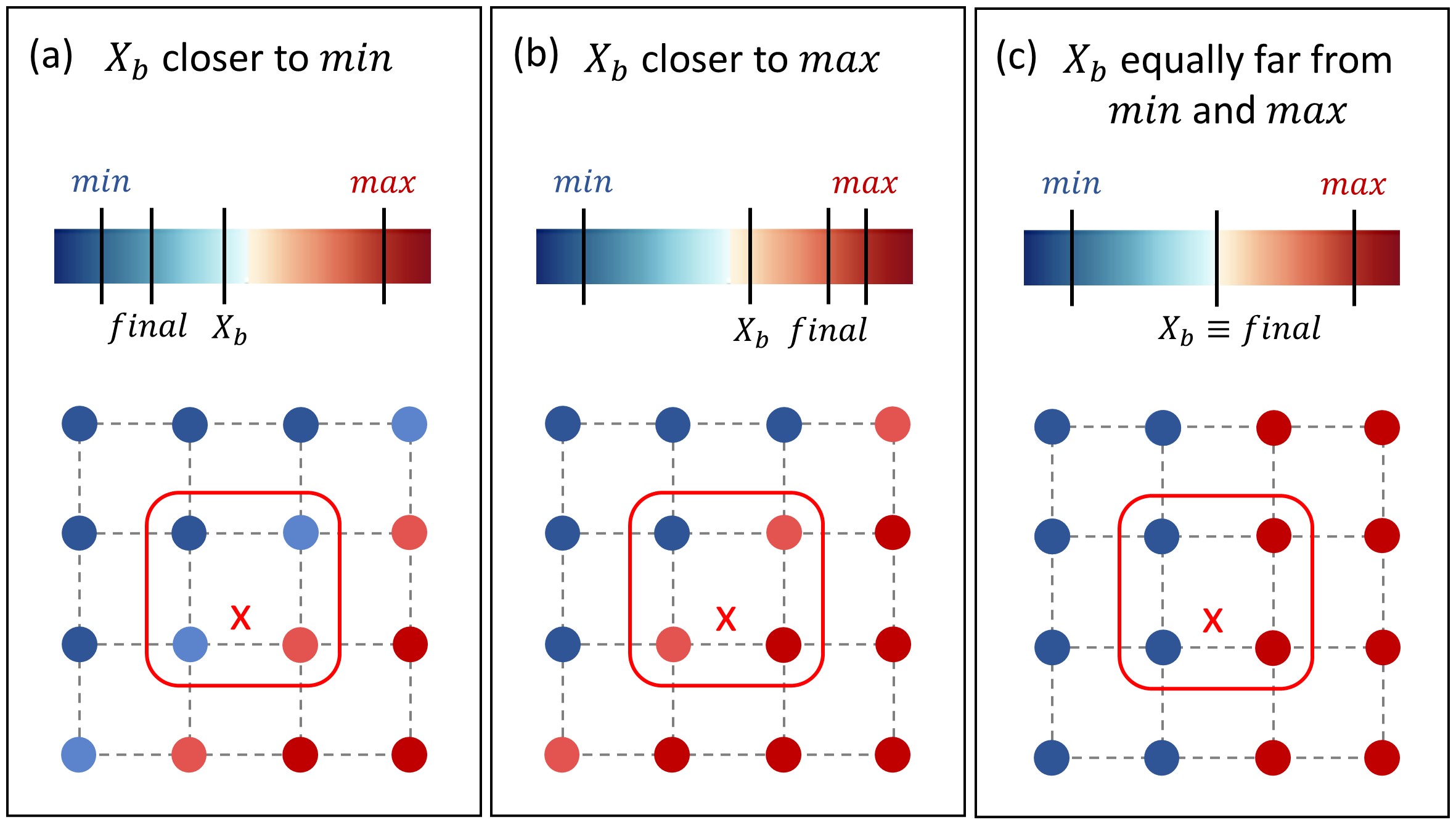}
\vspace{-6mm}
\caption{\label{fig:FIG3}{The three possible scenarios for interpolation, each illustrated through a four by four patch with the colour of the grid points indicating their intensity. The position of the interpolated grid point is shown by the cross and the four grid points used for the interpolation are the circled ones. If the bilinearly interpolated value $X_b$ is closer to the minimum, the interpolation favours lower intensities, and the final value is even closer to the minimum (a).  If $X_b$ is closer to the maximum, higher intensities are favoured and the final value is even closer to the maximum (b). If $X_b$ is equally far from minimum and maximum, the interpolated grid point is given this value (c).}}
\vspace{-4mm}
\end{figure}

If the interpolated point is far from any edges in the original image, it is assigned $X_b$ as its final value; otherwise, $X_b$ is used to guide the decisions taken in the next steps. More specifically, if the interpolated point is close to any detected edges, the algorithm considers the four by four patch that surrounds it in the original image and determines the maximum and minimum intensities in that region. Subsequently, the intensity of the interpolated grid point is determined through Algorithm \ref{alg:EdgePreserving} (EdgePreserving). Depending on the value of $X_b$, we have three possible scenarios:

\begin{enumerate}[nolistsep, topsep=0pt]
  \setlength{\itemsep}{0pt}
  \setlength{\parskip}{0pt}
  \setlength{\parsep}{0pt}
  
  \item $X_b$ is closer to the minimum intensity in the patch, so the algorithm favours the lower interpolated intensities (Fig.~\ref{fig:FIG3}\hyperlink{fig:FIG3}{(a)}). As shown in Algorithm \ref{alg:EdgePreserving}, whenever two intensities are interpolated (e.g., $br$ and $bl$, or $tr$ and $tl$), the computation is performed through Algorithm \ref{alg:2ValuesInterp} (2ValuesInterp). Among its inputs, Algorithm \ref{alg:2ValuesInterp} receives the two intensities under consideration and the weight that typically multiplies the lowest of the two in bilinear interpolation. It also receives a Boolean value, which in this scenario is \texttt{true} to indicate that the algorithm should favour the low intensity. This is done by increasing the interpolation weight by an amount that is proportional to the difference between the low intensity and the maximum in the patch.
  
  \item $X_b$ is closer to the maximum intensity in the patch, so the algorithm favours the higher interpolated intensities (Fig.~\ref{fig:FIG3}\hyperlink{fig:FIG3}{(b)}). Once again, Algorithm \ref{alg:2ValuesInterp} is used to interpolate groups of two intensities. However, in this case, 2ValuesInterp receives a \texttt{false} Boolean value as input, indicating that the algorithm should favour the high intensity. Therefore, the weight multiplying such intensity is increased by an amount proportional to the difference between this intensity and the minimum in the patch.
  
  \item $X_b$ is equally far from both minimum and maximum intensities (Fig.~\ref{fig:FIG3}\hyperlink{fig:FIG3}{(c)}), so this is the final value given to the grid point.
\end{enumerate} 
  Therefore, bilinear interpolation is only used if the interpolated value is equally far from both the minimum and maximum. However, to avoid generating very sharp edges, it is possible to modify the algorithm so that this happens for a range of values close to the centre of the interval bounded by the minimum and maximum (rather than at a single value). This is particularly useful at low frequencies, where it reduces staircasing artefacts, which would damage the reconstruction by generating non-physical effects.

\begin{algorithm}[H]
\vspace{3mm}
\caption{EdgePreserving}
\label{alg:EdgePreserving}
\begin{algorithmic}[1]
\footnotesize
\REQUIRE{The original image $IM$, the minimum and maximum intensities in the patch ($min_{patch}$ and $max_{patch}$), $X_b$, the distances $h$ and $v$, the intensities of the interpolated grid points ($br$, $bl$, $tr$, $tl$), and a vector $dist$ containing the factors that multiply $br$, $bl$, $tr$, $tl$, $b$ and $t$ in Algorithm \hyperlink{algo:bilinear}{1}
}
\AlgBlankLine
\ENSURE{The final value for the grid point}
\AlgBlankLine
\STATE $b_{1} \gets$ min($br$, $bl$)
\STATE $b_{2} \gets$ max($br$, $bl$)
\STATE $t_{1} \gets$ min($tr$, $tl$)
\STATE $t_{2} \gets$ max($tr$, $tl$)
\STATE $d_{1} \gets$ the factor that should multiply $b_{1}$ based on $dist$
\STATE $d_{2} \gets$ the factor that should multiply $t_{1}$ based on $dist$
\IF{abs($X_b-min_{patch}$) $<$ abs($X_b-max_{patch}$)}
\STATE $b \gets$ 2ValuesInterp ($IM$, $b_{1}$, $b_{2}$, $d_{1}$, $Low$=\texttt{True})
\STATE $t \gets$ 2ValuesInterp ($IM$, $t_{1}$, $t_{2}$, $d_{2}$, $Low$=\texttt{True})
\STATE $bt_{1} \gets$ min($b$, $t$)
\STATE $bt_{2} \gets$ max($b$, $t$)
\STATE $d_{3} \gets $ the factor that should multiply $bt_{1}$ based on $dist$
\STATE $output \gets$ 2ValuesInterp ($IM$, $bt1$, $bt2$, $d_{3}$, $Low$=\texttt{True})
\ELSIF{abs($X_b-min_{patch}$) $>$ abs($X_b-max_{patch}$)}
\STATE $b \gets$ 2ValuesInterp ($IM$, $b_{1}$, $b_{2}$, $d_{1}$, $Low$=\texttt{False})
\STATE $t \gets$ 2ValuesInterp ($IM$, $t_{1}$, $t_{2}$, $d_{2}$, $Low$=\texttt{False})
\STATE $bt_{1} \gets$ min($b$, $t$)
\STATE $bt_{2} \gets$ max($b$, $t$)
\STATE $d_{3} \gets $ the factor that should multiply $bt_{1}$ based on $dist$
\STATE $output \gets$ 2ValuesInterp($IM$, $bt_{1}$, $bt_{2}$, $d_{3}$, $Low$=\texttt{False})
\ELSE
\STATE $output \gets $ BilinearInterpolation($br$, $bl$, $tr$, $tl$, $h$, $v$)
\ENDIF
\AlgBlankLine
\end{algorithmic}
\end{algorithm}

\begin{algorithm}[H]
\caption{2ValuesInterp}
\label{alg:2ValuesInterp}
\begin{algorithmic}[1]
\footnotesize
\REQUIRE {The original image $IM$, the lowest of the two interpolated intensities ($a$), the highest of the two ($b$), the factor $r$ that typically multiplies intensity $a$ in bilinear intepolation (Algorithm \hyperlink{algo:bilinear}{1}), and a Boolean value $Low$ being \texttt{true} if the algorithm should favour low intensities and \texttt{false} otherwise}
\AlgBlankLine
\ENSURE{The value obtained by interpolating $a$ and $b$}
\AlgBlankLine
\STATE $MIN \gets$ min($IM$)
\STATE $MAX \gets$ max($IM$)
\IF{$Low$ is \texttt{True}}
\STATE $increase\gets \frac{MAX-a}{MAX-MIN}$
\STATE $w\gets r+increase*(1-r)$
\STATE $output\gets w*a+(1-w)*b$
\ELSE
\STATE $increase\gets \frac{b-MIN}{MAX-MIN}$
\STATE $w\gets (1-r)+increase*(r)$
\STATE $output\gets (1-w)*a+w*b$
\ENDIF
\AlgBlankLine
\end{algorithmic}
\end{algorithm}

\vspace{10mm}

\subsection{\label{subsec:wavefield_compr} Wavefield Compression}
\subsubsection{Ensuring convergence\label{subsubsec:convergence}}

To further reduce memory consumption, we rely on lossy compression of the acoustic wavefield, the solution of the forward wave equation. Using lossy compression to reduce the size of the temporal snapshots is possible because the wavefield exhibits sparsity in some domains. In other words, most of the wavefield information is represented by a small number of high-magnitude values, with the remaining values being zero or negligibly small in relative terms. However, lossy compression always introduces a certain amount of error, and this must be limited to ensure convergence of the minimization problem. 

In the next paragraphs, we recall the computations typically involved in FWI minimization and introduce the conditions needed to ensure convergence when the acoustic wavefield is affected by sources of error.

Line-search methods aim to iteratively improve a certain model $m$. At each iteration $k$, the update to the model $m_{k}$ is given by $\alpha_{k}s_{k}$, where $\alpha_{k}$ is the step length and $s_{k}$ is the search direction. $s_{k}$ is found by minimising the misfit functional $f$, which involves computing $\nabla f_{k}$, that is, the gradient of $f$ evaluated at $m_k$. To obtain $\nabla f_{k}$, it is necessary to calculate the Fréchet derivatives of $f$ with respect to $m$. This is typically done through the adjoint-state method, which allows to write the Fréchet derivatives, e.g. for the acoustic case, in the form \citep{operto},
\begin{eqnarray} 
K(x)= \int_{0}^{T} (D^2p)(x,t) \cdot (p^\dag)(x,t)\: dt \;,
\label{eq:frechet}
\end{eqnarray}
\noindent
where $p$ is the forward wavefield, $p^\dag$ is the adjoint wavefield and $D^2$ is a second-order temporal differential operator. Our aim is to compress $D^2p$ during the forward run and decompress it during the adjoint run. Since the decompressed wavefield $\tilde{D^2p}$ is approximate and is used in place of $D^2p$ in Eq. \autoref{eq:frechet}, the resulting gradient is also an approximation. 

To ensure convergence to a minimum, the approximate gradient $\tilde{g}$ must be sufficiently close to the exact gradient $g$. More precisely, convergence can be guaranteed if the inexactly computed search directions $\tilde{s}_k$ satisfy the so-called angle condition\citep{nocedal}, that is,
\begin{eqnarray} 
(g_k,\tilde{s}_k) \le -\beta \: ||g_k|| \cdot ||\tilde{s}_k||
\label{eq:angcond}
\end{eqnarray}
\noindent
for some $\beta > 0$, at all iterations $k$. Since the ratio $(g_k,\tilde{s}_k) \,/ \, (||g_k|| \cdot ||\tilde{s}_k||)$ represents the cosine of the angle between the inexactly computed search direction $\tilde{s}_k$ and the exact gradient $g_k$, Equation \autoref{eq:angcond} implies that such angle must be strictly smaller than 90° in order to ensure convergence. Because in the steepest descent method $\tilde{s}_k = \tilde{g}_k$, such condition sets a limit to the angle between the approximate gradient and the exact one. Intuitively, we expect the angle between the two gradients to depend on the error introduced when compressing the forward wavefield. Therefore, we can control such angle by setting a limit to the relative error allowed between the original and decompressed wavefields.

\subsubsection{Compression techniques\label{subsubsec:compr_tech}}

For reducing the size of the acoustic wavefield, we combine temporal and spatial downsampling with other lossy compression techniques. The use of spatial and temporal downsampling is motivated by the fact that the maximum spatial and temporal steps respectively obtained through Eq.\autoref{eq:dx_max} and Eq.\autoref{eq:cfl} are much smaller than those prescribed by the Nyquist-Shannon theorem. Based on this, we only store the wavefield every $k$ time steps, where $k$ is such that the resulting temporal sampling rate satifies the condition set by the Nyquist-Shannon theorem. In the rest of the paper, we use $k$ to quantify the amount of temporal downsampling and refer to it as temporal downsampling ratio (TR). At each of the stored time steps, the wavefield is downsampled in space through bicubic interpolation. An edge-preserving interpolation is not necessary in this case because, contrary to the recovered acoustic speeds, the acoustic pressure varies gradually through the domain and no high-contrast areas are present. The spatial downsampling ratio (SR), that is, the ratio between the dimensions of the original grid and the new one, was determined through experimentation and was such that the resulting grid spacing would always be smaller than the limit set by Nyquist-Shannon theorem.

Subsequently, we further compress the wavefield by representing it in an appropriate domain, eliminating low-information values through thresholding, and then storing the resulting values in a sparse, half-precision representation. We investigated three domains in which to apply our techniques: the standard spatial domain, the wavelet domain, and the wave atom domain. For computing the wavelet transform, we rely on the Daubechies db5 wavelet because we experimentally determined that, in this case,  it leads to better results when compared to other types of wavelets. As for the wave atoms, these are a family of wave packets that provide an optimally sparse representation of images with oscillatory patterns and, therefore, are typically well suited for compressing wave-equation solutions \citep{wa}. Different variants of wave atoms exist. Here, we use the orthobasis variant because it allows to reduce redundancy and is therefore better suited for compression. 

The proposed sparsity-based compression is performed as follows. First, a hard thresholding approach is used to identify and store the values of the wavefield (or wavelet/wave atom coefficients) containing most information and discard the remaining ones. As previously mentioned, thresholding is used to enforce a sparse representation of the wavefield. The value of this threshold is identified based on the amount of compression error allowed on the (spatially downsampled) wavefield. More specifically, we rely on the mean of the n-highest point-wise relative errors, calculated as the difference between the original and decompressed wavefields, normalised by the dynamic range of the original one. We heuristically determine the value $n = 15$ as that which better describes the compression performance across the different domains studied and for the models used in this study. We refer to this value of the error as $\epsilon_{rel}$.

\begin{figure}
\includegraphics[width=\reprintcolumnwidth]{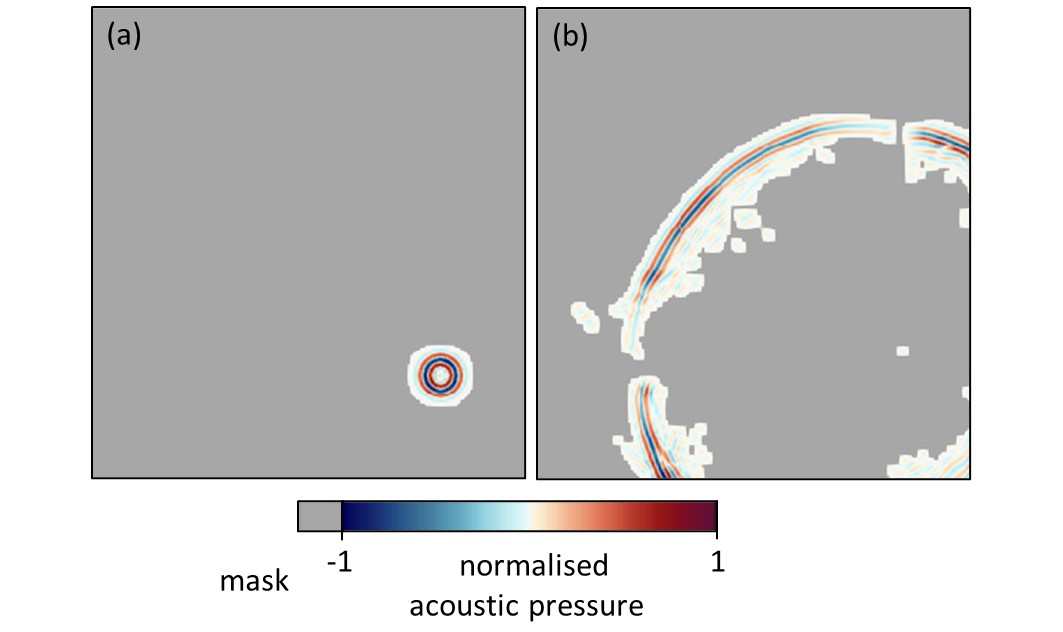}
\vspace{-6mm}
\caption{\label{fig:FIG4}{An example wavefield at two time steps of the forward run, with the discarded parts indicated in grey. The time steps approximately correspond to 4\% (a) and 31\% of the overall runtime. In both cases, the wavefield has been normalised by the maximum absolute value appearing in it.}}
\vspace{-3mm}
\end{figure}

Having identified the threshold, all values below it are discarded, while the values above it are stored in a sparse representation. Figure \ref{fig:FIG4} shows an example wavefield at different time steps, with the discarded part highlighted in grey. Once the most important values have been identified and saved, these are requantised and stored using 16-bit floating point precision. 

During the adjoint run, the decompression takes place, firstly, by undoing the requantisation to retrieve the values in 32-bit floating point precision and then by recasting the sparse representation of the wavefield into its dense counterpart. At this point, if operations were performed in the wavelet or wave atom domains, the spatial wavefield values are recovered by computing the inverse transform. Finally, the wavefield is upsampled to its original size.

Moreover, linear interpolation is used during the adjoint run to obtain an approximation of the states discarded by temporal downsampling. Although in this study we chose linear interpolation for its simplicity and low computational overhead, other types of interpolation might be used, such as spline interpolation, depending on the requirements of the application at hand.

\subsection{\label{subsec:experiments} Numerical experiments}

\begin{figure}[ht]
\hypertarget{fig:FIG5}{}
\includegraphics[width=\reprintcolumnwidth]{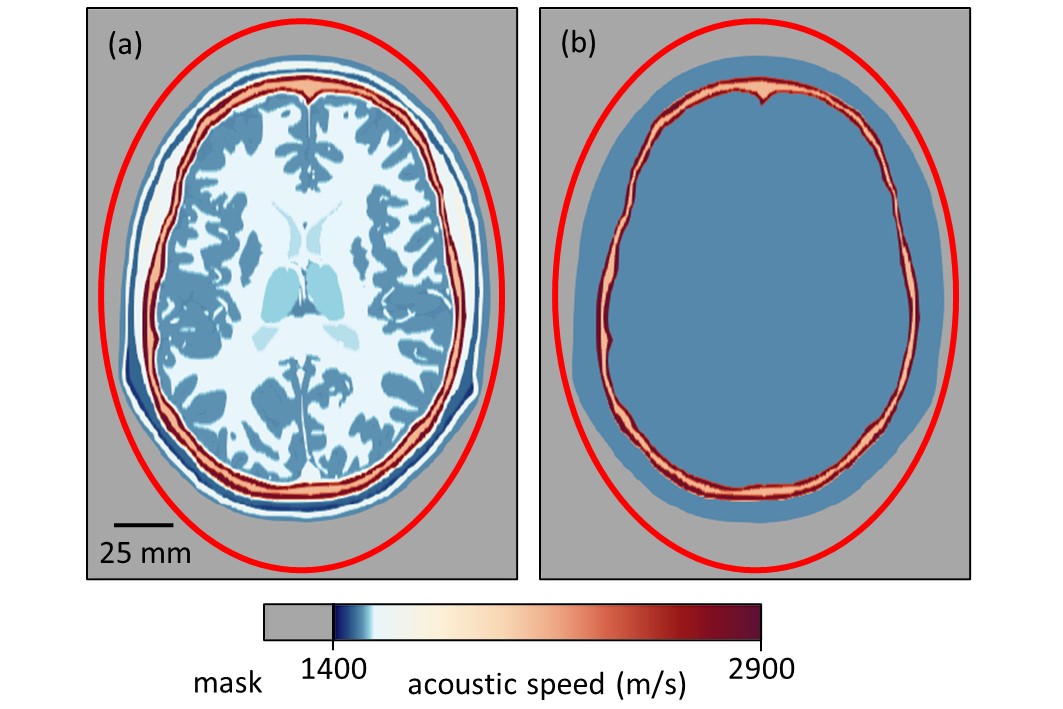}
\vspace{-5mm}
\caption{\label{fig:FIG5}{The model being imaged and the imaging setup for all numerical experiments. The set of synthetic observed data used in all tests was generated by solving the wave equation for an $in$  $silico$ model of the acoustic speed in a human head (a). In each experiment, the starting model consisted of the skull located in a homogeneous medium with the acoustic speed of water (b). The ellipses in the figure indicate the location of the 2D array of transducers used for generating the data and inverting it.}}
\vspace{-3mm}
\end{figure}

In order to test the proposed multi-grid and compression methods, we applied them to image a numerical model of the human head. The model being imaged can be seen in Fig.~\ref{fig:FIG5}\hyperlink{fig:FIG5}{(a)}, with the ellipse showing the location of the 120 point transducers used as sources and receivers. This model was obtained from the segmented MIDA model\citep{mida}, for which acoustic properties were assigned according to experimental measurements available in the literature, as seen in Ref. \citen{lluis}. From this 3D model, a single 2D slice was taken for all experiments presented here. In all cases, the starting model consisted of the true skull, located within a homogeneous medium with the acoustic speed of water (Fig.~\ref{fig:FIG5}\hyperlink{fig:FIG5}{(b)}).

Imaging was performed using a Ricker wavelet with a centre frequency of 300 kHz. The inversion was carried out in five frequency bands from 200 to 600 kHz, using 10 iterations per band. At each iteration, a subset of 12 sources was selected. For every source, 200 $\upmu$s of data were generated with a time spacing of 0.08 $\upmu$s, as determined by the CFL condition (Equation \autoref{eq:cfl}, with $\mu = 0.55$). For each frequency band, the grid sampling was calculated through Eq. \autoref{eq:dx_max}, using $n = 5$. 

For all the experiments, the 2D acoustic wave equation with homogeneous density was solved using a time-domain, finite-difference method implemented in Devito, a domain-specific language for the automatic generation of FD code \citep{devito}. The FD stencils used were eleventh-order accurate in space and fourth-order accurate in time. 

Although we relied on 2D FWI in this study, all the techniques presented here are directly applicable to 3D. Of the proposed approaches, the only algorithm that would need to be adapted for 3D FWI is the edge-preserving interpolation, which could be done trivially.

\subsection{\label{subsec:measures} Measures for assessing the results}

To assess the efficacy of our methods, we rely on several measures. Among these, the mean overall compression factor (CF) is used to quantify the amount of memory saved (on average) during the simulation. This is a mean, across all iterations, of the ratio between the size of the original wavefield (i.e., its size in conventional FWI) and the size of the compressed one. 

The quality of the inexactly computed gradients (obtained with the compressed wavefields) is determined by comparing them to the exact gradients using two criteria. Firstly, we use the angular difference $\uptheta$, which represents the angle between the exact gradient $g$ and the inexact gradient $\tilde g$, and is calculated as,
\begin{eqnarray} 
\cos{\theta} = \frac{(\tilde g,g)}{||\tilde g|| \cdot ||g||} \;\;.
\label{eq:angdiff}
\end{eqnarray}

A smaller angular difference means that the model update resulting from the inexact gradient will be similar to the update obtained with the exact one. Secondly, we quantify the similarity between exact and inexact gradients through the structural similarity index (SSIM), which varies from 0 to 1, with 1 indicating maximum similarity\citep{ssim}. For both the angular difference and SSIM, we show mean values, computed across the whole inversion process. 

For some of our numerical tests we also include the overhead in simulation time due to compression and decompression (OV) with respect to multi-grid FWI without compression, as well as the range of values taken by the instantaneous compression factor (ICF), i.e. the CF calculated at a single time step.

\section{\label{sec:results} Results}

The multi-grid approach allowed us to reduce the reconstruction time of FWI by approximately 30\% for the tested model. This is due to the fact that, while in conventional FWI the inversion takes the same amount of time for each frequency band, in multi-grid FWI the inversion of the lowest frequencies takes a shorter time due to the reduced amount of computations associated with coarser grids. For the model imaged in this study, such reduction in computational time varies from 64\% for the first frequency band to 0\% for the last band. 

Because coarser grids require less memory to be stored, the multi-grid approach also lowers memory consumption, by an amount similar to the reduction in computational time. A more significant decrease in memory usage was obtained by combining the multi-grid approach with lossy compression. 

\begin{figure}[b]
\vspace{1mm}
\hypertarget{fig:FIG6}{}
\includegraphics[width=\reprintcolumnwidth]{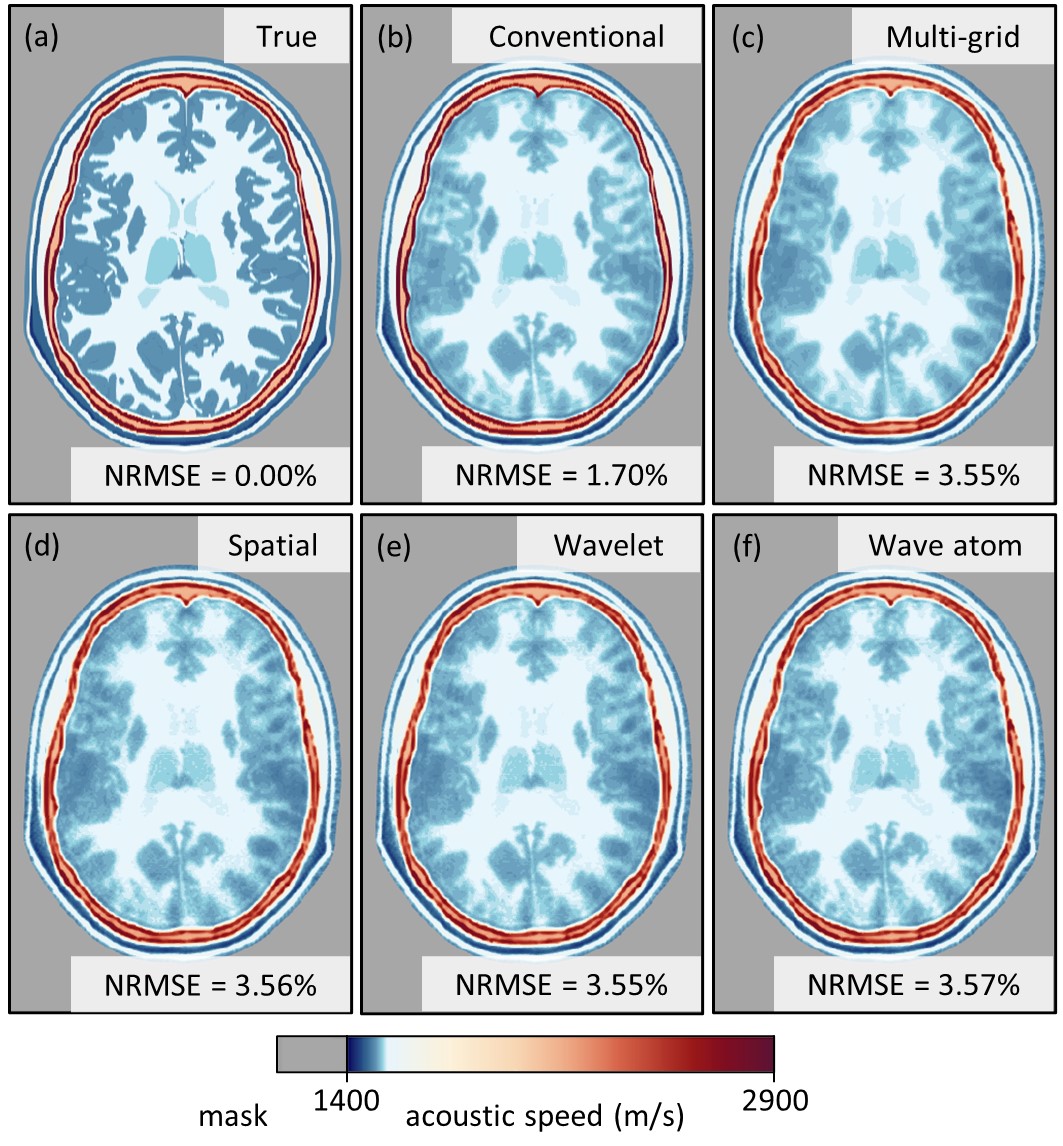}
\vspace{-6mm}
\caption{\label{fig:FIG6}{Impact of our methods on the \textit{in silico} FWI reconstruction. When conventional FWI is used (b), the recovered model closely matches the true model (a). When multi-grid FWI is used (c), an accurate reconstruction is obtained, but the recovered speed of the brain soft tissue is slightly lower than it should be. When compression is used jointly with the multi-grid approach, the impact is negligible, independently of whether the sparsity-based approach is applied in the spatial (d), wavelet (e) or wave atom domain (f). For each reconstruction, the figure also shows the NRMSE with respect to the true model.}}
\vspace{-5mm}
\end{figure}

As explained in Section \ref{subsubsec:compr_tech}, we tested the performance of our compression techniques in three domains. In all cases, we set the temporal downsampling ratio, spatial downsampling ratio and relative error $\epsilon_{rel}$ to the values that ensure the highest compression while negligibly affecting the accuracy of the reconstruction (10, 2.2 and 9\%, respectively). In Fig.~\ref{fig:FIG6}, we show the recovered model obtained in each case, together with the true model, the model recovered by conventional FWI, and the one obtained using the multi-grid approach only. For each reconstruction, the normalised root-mean-square error (NRMSE) with respect to the true model is included in the figure.

\begin{figure}[ht]
\vspace{1mm}
\hypertarget{fig:FIG7}{}
\includegraphics[width=\reprintcolumnwidth]{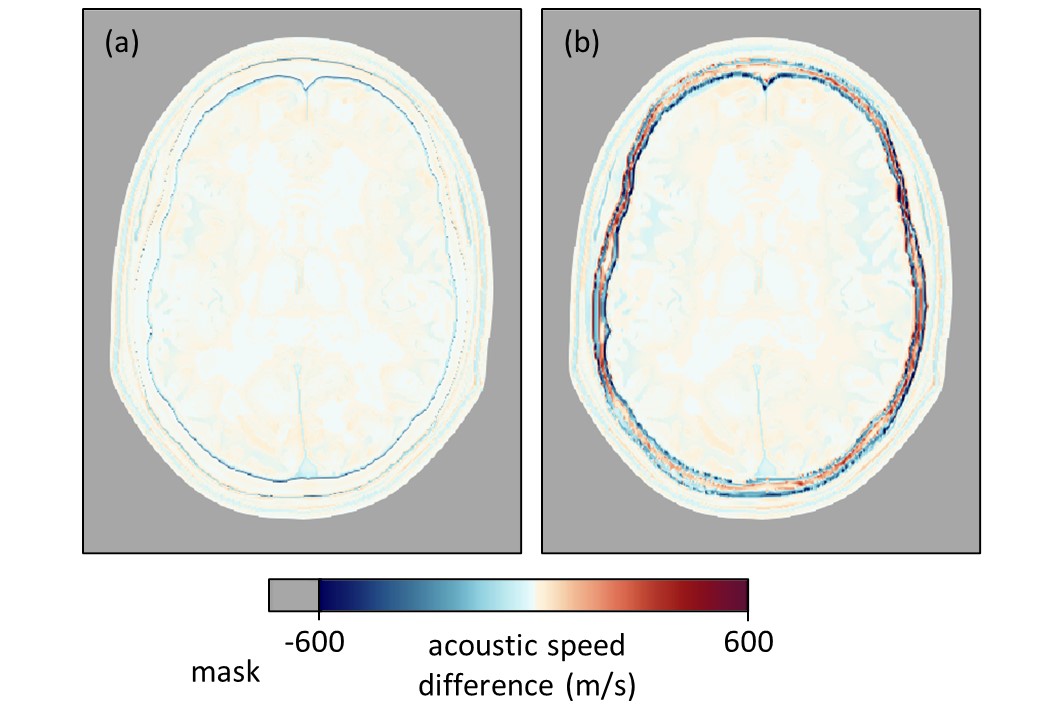}
\vspace{-5mm}
\caption{\label{fig:FIG7}{Difference maps for conventional and multi-grid FWI, obtained by subtracting the true model from the recovered one. When conventional FWI is used (a), the error is homogeneously distributed, whereas, when multi-grid FWI is used (b), the error is concentrated at the skull, which is damaged by the resampling.}}
\vspace{-3mm}
\end{figure}

As can be seen, the models obtained through multi-grid FWI with compression (Figs.~\ref{fig:FIG6}\hyperlink{fig:FIG6}{(d)-}\ref{fig:FIG6}\hyperlink{fig:FIG6}{(f)}) are in good agreement with the one recovered by conventional FWI (Fig.~\ref{fig:FIG6}\hyperlink{fig:FIG6}{(b)}). Moreover, the models recovered by relying on compression do not present significant differences with respect to the one obtained using the multi-grid approach only (Fig.~\ref{fig:FIG6}\hyperlink{fig:FIG6}{(c)}), both qualitatively and quantitatively. However, in all reconstructions obtained through multi-grid FWI (with or without compression), the NRMSE is approximately double the one of the model recovered by conventional FWI. To visualise how such error is distributed, Figure \ref{fig:FIG7} shows difference maps between the true model and the models obtained using conventional FWI (Fig.~\ref{fig:FIG7}\hyperlink{fig:FIG7}{(a)}) and multi-grid FWI without compression (Fig.~\ref{fig:FIG7}\hyperlink{fig:FIG7}{(b)}). As can be deduced from the figure, in the model recovered by multi-grid FWI the error is concentrated in the skull region, which has been damaged as a result of resampling the model multiple times. This damage to the skull leads to an angular difference of 28.7° between exact and approximate gradients in the multi-grid inversion without compression, which in turn influences the recovered speeds of the intracranial soft tissue. In fact, these are slightly lower in the model recovered by multi-grid FWI with respect to the one obtained with conventional FWI, especially in the areas close to the skull. However, as evidenced from the recovered models in Fig.~\ref{fig:FIG6}, this effect does not prevent a successful reconstruction.

As previously mentioned, introducing lossy compression further reduced the memory consumption. In Table \ref{tab:table1}, we show the mean compression factor and angular difference obtained through the experiments relying on multi-grid FWI with compression, for each domain we tested. For the tests where compression is performed in the original spatial domain and the wavelet domain, we also report the computational overhead. For the experiment relying on wave atoms, this measure is not presented here because the value obtained was not representative of the real performance of the wave atom transform. This is because, in order to ensure compatibility with existing inversion codes,  the transform was re-written in Python for this study, based on the original MATLAB code \citep{wa}, and has not yet been optimised.

\begin{table}[t]
\caption{\label{tab:table1}Comparison of the results obtained by applying our compression techniques in the different domains considered. }
\setlength{\tabcolsep}{17.5pt}
\begin{adjustbox}{width=\reprintcolumnwidth}
\begin{tabular}{c|ccc}
\hline\hline
 & CF & $\uptheta$ ($^{\circ}$)   & OV (\%) \\
\hline
Spatial & 3595 & 37.4 & 1.51 \\
Wavelet & 3905 & 37.0 & 3.68 \\
Wave atom & 2559 & 37.9 & - \\
\hline\hline
\end{tabular}
\end{adjustbox}
\vspace{2mm}

\end{table}

From Table \ref{tab:table1}, we can see that the highest compression is achieved when our techniques are applied in the wavelet domain, while a slightly lower CF is obtained when they are applied directly on the spatial-domain wavefield. This suggests that, although the wavefield itself is sparse in the spatial domain, its representation in the wavelet domain is even sparser. A lower CF is obtained when compression is performed in the wave atom domain. As would be expected, the mean angular difference is comparable in all cases, with no significant differences between them. As for the computational overhead, this is negligible in the two cases compared, even if significantly higher for the wavelets than for the unaltered wavefield.

\begin{figure}[ht]
\hypertarget{fig:FIG8}{}
\includegraphics[width=\reprintcolumnwidth]{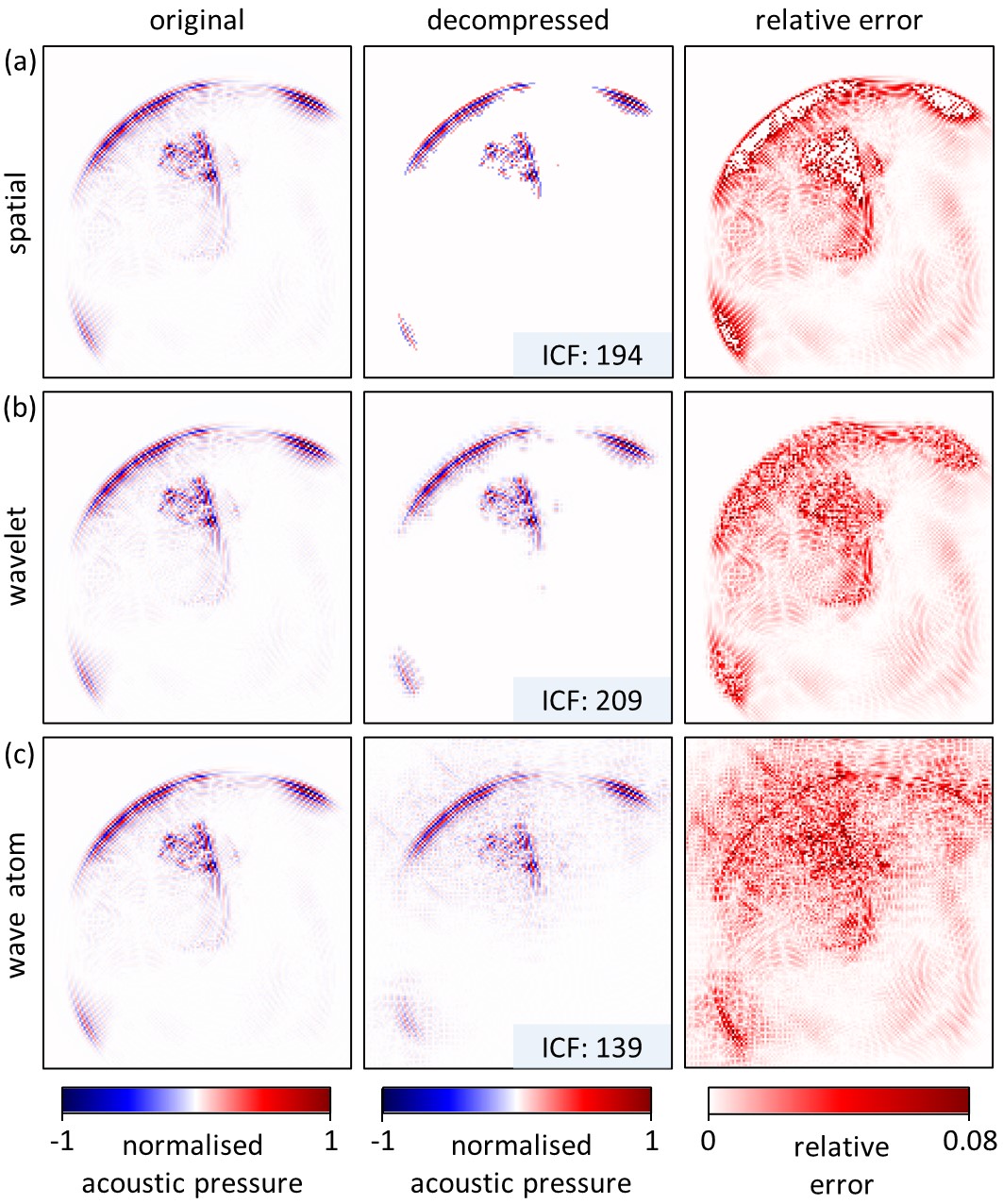}
\vspace{-5mm}
\caption{\label{fig:FIG8}{The effect of sparsity-based compression on the wavefield, for each of the tested domains. These include the original spatial domain (a), the wavelet domain (b) and the wave atom domain (c). In all cases, the left panel shows the original wavefield, the central panel shows the decompressed one and the right panel shows the relative error between the two. All wavefields have been normalised by the maximum magnitude in the original one. ICF indicates the compression factor achieved on the wavefield at the time step being shown and does not include the contribution of the multi-grid approach.}}
\vspace{-3mm}
\end{figure}

To better understand these results, we show in Fig.~\ref{fig:FIG8} the effects of sparsity-based compression on an example wavefield (at a particular time step), for each of the tested domains. The figure includes the spatially downsampled wavefield before compression, the decompressed wavefield (before upsampling), and the map of the relative error between the two. For each domain, the figure also shows the instantaneous compression factor achieved on the wavefield (at the time step being shown). These ICF values do not include the contribution of the multi-grid approach, as this varies based on the frequency band.

By comparing the decompressed wavefields in Figs.~\ref{fig:FIG8}\hyperlink{fig:FIG8}{(a)} and \ref{fig:FIG8}\hyperlink{fig:FIG8}{(b)}, we can see that applying our techniques in the wavelet domain rather than in the spatial one allows us to retain a larger part of the original wavefield, while also achieving a higher ICF. This again emphasizes that the wavefield is sparser in the wavelet domain. From Fig.~\ref{fig:FIG8}\hyperlink{fig:FIG8}{(c)}, it is possible to see that, when the wave atom transform is used, a large portion of the original wavefield is retained, but the decompressed wavefield is corrupted by artefacts. This results from the loss of information caused by the hard-thresholding approach, as the artefacts become more evident when less coefficients are stored. Moreover, thresholding has a more important impact on the largest pressure values compared to when the wavelet transform is used, which limits the compression level that can be achieved with wave atoms. Because our purpose is to reduce memory use as much as possible while also preserving the most important information, the wavelet and spatial domains represent more suitable alternatives than the wave atom one for the case tested in this study. Therefore, for both these domains we performed additional experiments, which are subsequently presented. 

To determine how each technique contributes to the overall CF, we also ran some experiments where compression is performed only through temporal and spatial downsampling. The results of these experiments are shown in Table \ref{tab:table2}. Table \ref{tab:table3}, instead, shows the results of the tests that relied on all the proposed techniques, with sparsity-based compression performed either in the spatial (experiment 5 to 7) or the wavelet (experiment 8 to 10) domain, for different levels of $\epsilon_{rel}$. The table also includes the experiments that resulted in the recovered models shown in Figs.~\ref{fig:FIG6}\hyperlink{fig:FIG6}{(d)} and \ref{fig:FIG6}\hyperlink{fig:FIG6}{(e)} (experiments 7 and 10 respectively).

\begin{table}[ht]
\caption{\label{tab:table2}Results for the experiments performed without sparsity-based compression, for various configurations of temporal and spatial downsampling ratios. Such settings are specified in columns 2 and 3, while columns 4-8 refer to the results (calculated as described in Section \ref{subsec:measures}).}
\setlength{\tabcolsep}{6.5pt}
\begin{adjustbox}{width=\reprintcolumnwidth}
\centering
\begin{tabular}{c|ccccccc}
\hline\hline
 & TR & SR & CF & $\uptheta$ ($^{\circ}$) & SSIM & OV (\%) & ICF \\
\hline
Exp. 1 & 5 & 1 & 9 & 29.4  & 0.74 & 0.89 & 1 - 3\\
Exp. 2 & 10 & 1 & 17 & 28.9 & 0.73 & 0.48 & 1 - 3\\
Exp. 3 & 10 & 2 & 69 & 32.4 & 0.74 & 1.05 & 4 - 11\\
Exp. 4 & 10 & 2.2 & 84 & 33.8 & 0.73 & 1.02 & 5 - 14\\
\hline\hline
\end{tabular}
\end{adjustbox}
\vspace{2mm}
\end{table}

\begin{table*}[ht]
\caption{\label{tab:table3}Results for the experiments that relied on all the proposed techniques, including sparsity-based compression, performed either in the spatial domain (experiments 5 to 7) or in the wavelet domain (experiments 8 to 10). Columns 3-5 refer to the settings for each experiment (defined in Section \ref{subsubsec:compr_tech}), while columns 6-10 refer to the results.}
\setlength{\tabcolsep}{15pt} 
\begin{adjustbox}{width=\textwidth}
\centering
\begin{tabular}{ c | c | c c c c c c c c }
\hline\hline
  &  &  TR  &  SR  &  $\epsilon_{rel}$ (\%)  &  CF  &  $\uptheta$ ($^{\circ}$)  &  SSIM & OV (\%)  & ICF \\
\hline
\multirow{3}{*}{Spatial} & Exp. 5 & 10 & 2.2 & 3 & 1786  & 36.2 & 0.69 & 1.55 & 43 - 71237 \\
& Exp. 6 & 10 & 2.2 & 6 & 2728 & 36.7 & 0.68 & 1.52 & 84 - 71237 \\
& Exp. 7 & 10 & 2.2 & 9 & 3595 & 37.4 & 0.68 & 1.51 & 101 - 71237 \\
\hline
\multirow{3}{*}{Wavelet} & Exp. 8 & 10 & 2.2 & 3 & 1832 & 33.5 & 0.70 & 3.72 & 45 - 85485 \\
& Exp. 9 & 10 & 2.2 & 6 & 3189 & 35.1 & 0.69 & 3.72 & 100 - 85485 \\
& Exp. 10 & 10 & 2.2 & 9 & 3905 & 37.0 & 0.68 & 3.68 & 124 - 85485 \\
\hline\hline
\end{tabular}
\end{adjustbox}
\end{table*}

The results in Table \ref{tab:table2} and Table \ref{tab:table3} allow us to identify the techniques that contributed most to the obtained compression levels. From Table \ref{tab:table2}, we see that, as expected, the CF follows a linear relationship with the temporal downsampling ratio and a quadratic relationship with the spatial downsampling ratio (due to the 2D nature of the experiment). Since the temporal and spatial downsampling are limited by the Nyquist-Shannon theorem, the amount of compression that these techniques can achieve is necessarily limited. On the other hand, if the temporal and spatial downsampling ratios are kept constant, introducing sparsity-based compression increases the CF by a factor of up to 46 (as can be seen from Table \ref{tab:table3}). Therefore, most of the compression is due to this approach.

\begin{figure}[h]
\hypertarget{fig:FIG9}{}
\includegraphics[width=\reprintcolumnwidth]{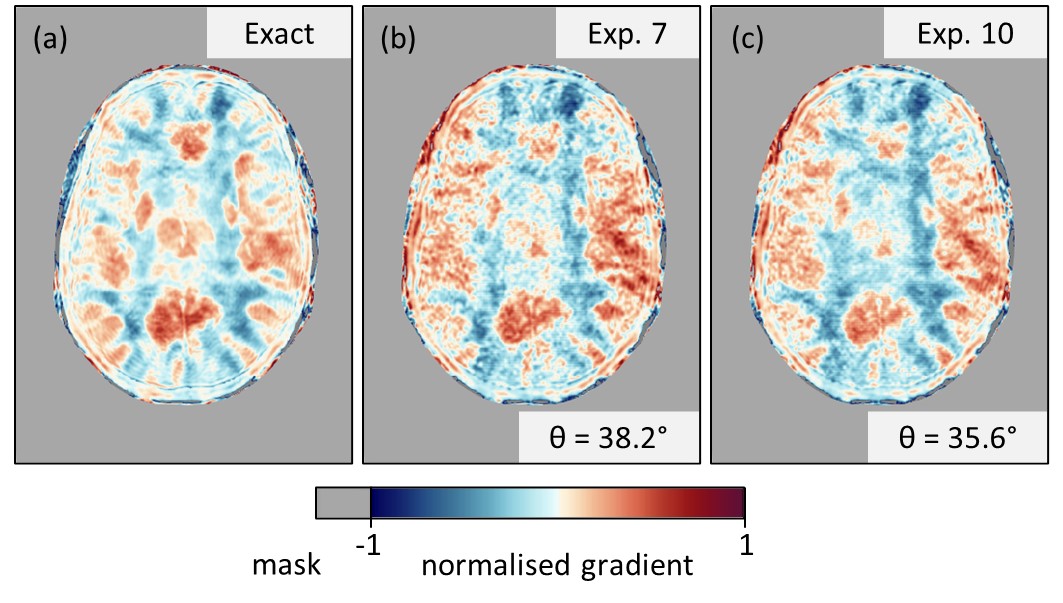}
\vspace{-6mm} 
\caption{\label{fig:FIG9}{The impact of the multi-grid approach and lossy compression on the gradients. The gradients shown here have been extracted at iteration 25 of a conventional FWI inversion (a), experiment 7 (b) and experiment 10 (c). To facilitate the comparison, the gradients have been normalised with respect to each other. For each gradient, $\uptheta$ indicates the angular difference with respect to the exact one.}}
\vspace{-3mm}
\end{figure}

Table \ref{tab:table3} also shows that, given a certain amount of error allowed on the wavefield, applying our techniques in the wavelet domain rather than on the wavefield itself always results in higher compression factors with similar levels of angular difference between exact and approximate gradients. As we might expect, independently of the domain in which compression is performed, the mean angular difference increases slightly when a larger amount of relative error is allowed. However, even with a mean CF of 3905, the inexactly computed gradient is still sufficiently similar to the exact one, with an angle of 37° between the two. To visualise how similar the gradients are when the angular difference between them is 37°, in Fig.~\ref{fig:FIG9} we show the gradients corresponding to iteration 25 for experiments 7 and 10, together with the exact gradient. It is possible to see that, although the inexact gradients exhibit some structural differences with respect to the exact one, they present similar feature distributions. Considering that in the experiments relying on multi-grid FWI without compression we obtained a mean angular difference of 28.7°, most of the angular difference is due to the multi-grid approach rather than lossy compression.

\begin{figure}[h]
\vspace{1mm}
\includegraphics[width=\reprintcolumnwidth]{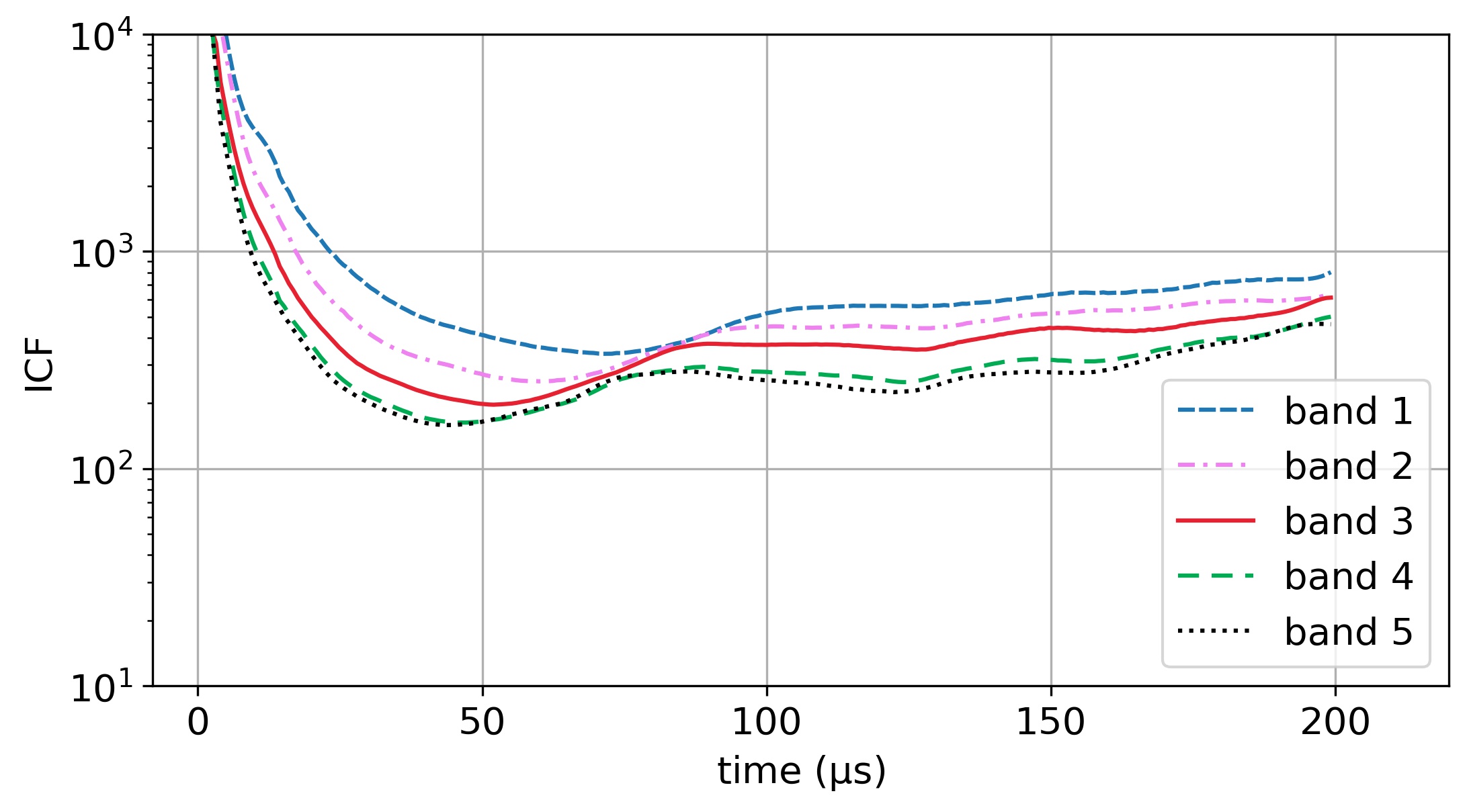}
\vspace{-6mm}
\caption{\label{fig:FIG10}{Instantaneous compression factor over time for five wavefields in experiment 10, one for each frequency band. All five wavefields are taken from the same source, during the fifth iteration of each band. The vertical axis has been truncated to a factor of 10$^4$; in the first few time steps, the ICF reaches values close to the highest one shown in Table \ref{tab:table3} for experiment 10 (85485).}}

\end{figure}

Finally, from Table \ref{tab:table3} we can also see that the instantaneous compression factor takes a large range of values. More specifically, the ICF is higher at the initial time steps of the simulation and progressively becomes lower. This is because, in the first few time steps, the wave occupies a small portion of the domain and most of the wavefield is discarded; as the wave expands, sparsity becomes lower and the compression level is reduced. This is illustrated in Fig.~\ref{fig:FIG10}, which shows how the ICF varies with time during a forward run for each frequency band. 

\section{\label{sec:discussion} Discussion}

The results presented here showcase that our methods are capable of reducing the reconstruction time of conventional FWI by approximately 30\% and its memory consumption by up to three orders of magnitude, while retaining the high accuracy of the reconstructions.

These results are of high importance for the clinical applicability of FWI because, in contexts like stroke imaging, time to treatment has a considerable impact on patient outcomes\citep{neuro2}. Moreover, as FWI is brought closer to clinical practice, graphics processing units (GPUs) could prove an important tool in accelerating the solution of the wave equation\citep{lluis, perez}. Since GPUs have limited working memory and the data transfer from/to host memory is computationally expensive, being able to reduce the wavefield size through compression could prove highly important when using these devices to accelerate computations.

As previously explained, the reduced reconstruction time is a result of the grid size changing based on the inverted frequencies. Since the computational cost, and therefore the reconstruction time, are proportional to the number of grid points, using coarser grids at low frequencies allows us to reduce both these measures. The multi-grid approach also contributes to lower memory use, although in smaller part with respect to lossy compression.

Our compression techniques achieve promising results both when they are applied on the wavefield itself, and its representation in the wavelet and wave atom domains. However, the accuracy and compression level obtained in the wavelet domain are generally higher than the others compared in this study, suggesting that the wavefield is sparser in this domain when imaging the head model used. Models with similar structural patterns to the one used here should lead to similar results, but it is important to note that significantly different models could exhibit different levels of sparsity in each of these domains.

Compared to the use of black-box lossy compressors like zfp\citep{kukreja1}, our methods allow much higher compression factors (up to two orders of magnitude), while still maintaining a low mean angular difference between the exact and inexact gradients. This is true independently of the domain in which our compression techniques are applied. As shown in Section \ref{sec:results}, such compression levels are mostly due to sparsity-based compression. This proves the benefit of exploiting our knowledge about the structure of the wavefields to tailor lossy compression to its application in FWI.

Furthermore, the proposed compression methods are applicable independently of the chosen numerical discretisation scheme. Therefore, while we relied on finite differences to carry out this study, the proposed techniques can be readily used for other numerical methods such as finite elements or pseudo-spectral methods. 

Another advantage of our lossy compression approaches is that they result in small computational overhead both when they are applied in the spatial domain and the wavelet domain. Even in the second case, where the use of the wavelet transform leads to additional computations, the computational overhead is only 3.7\% of the simulation time of multi-grid FWI without compression. This is significantly less than the overhead that characterises checkpointing methods, which generally require an amount of additional computations that is in the order of one forward simulation\citep{timerev}. As regards the computational overhead that results from applying our techniques in the wave atom domain, we found this value to be much higher than that obtained in the wavelet domain. However, as explained in Section \ref{sec:results}, this is mainly due to the fact that the wave atom transform was re-implemented in Python to guarantee compatibility with existing codes and has not been optimised yet. Consequently, this result is not an accurate reflection of the typical performance of the wave-atom transform (in terms of computation time).  

Regarding the angular difference between exact and approximate gradients, it is important to note that, while this is low enough to ensure convergence to some minimum, convergence to the global minimum is not guaranteed due to the non-convex nature of the problem. This can be mitigated by changing the compression threshold adaptively with the norm of the gradient, so as to gradually include more information toward the end of the simulation, as suggested in Ref. \citen{boehm}.

Based on our results, the biggest contributor to the angular difference between exact and approximate gradients is the multi-grid approach, with the different forms of compression having little influence on the values of this measure. As mentioned in Section \ref{sec:results}, this is due to the impact of the resampling process on high-contrast areas. Therefore, future work will explore possible improvements of the interpolation algorithm to reduce errors in these areas. 

The study presented here has focused on a 2D model due to its computational simplicity. However, realistic imaging applications will require 3D implementations. From the proposed techniques, the edge-preserving interpolation is the only one that would require some modifications in order to be used for 3D FWI. These modifications, however, would be straightforward and will be the focus of future research.

\section{\label{sec:conclusion} Conclusion}

Early diagnosis through brain imaging is crucial for the treatment of neurological diseases like stroke, which leads to a currently unmet need for fast, portable, and high-resolution neuroimaging. Full-waveform inversion (FWI) represents a promising modality that has the capacity to achieve this. However, the clinical applicability of FWI is limited by its high computational cost and memory requirements.

For this reason, we have developed a combination of techniques aimed at rendering FWI more computationally efficient. More specifically, we have exploited the fact that temporal frequencies are introduced gradually into FWI inversions to reduce computations and memory use through a frequency-adaptive spatial discretisation. Furthermore, we have combined this approach with multiple lossy compression techniques that take advantage of the sparsity of acoustic wavefields in different domains to further reduce their memory footprint. Numerical tests have shown that our methods can reduce memory consumption by up to three orders of magnitude and reconstruction time by 30\%, with negligible impact on the quality of the recovered model. 

The methodology introduced here will have a significant impact in the deployment of FWI in clinical scenarios, where faster image reconstructions can lead to saved lives and improved patient outcomes. Additionally, the results presented could have broader applicability beyond brain imaging, in geophysical FWI, but also in other physics-constrained optimisation problems in fields such as aeronautics and non-destructive testing.

\section*{\label{sec:acknowledgments} Acknowledgments}

The work of Carlos Cueto was supported by the Engineering and Physical Sciences Research Council (EPSRC) Centre for Doctoral Training in Medical Imaging under Grant EP/L015226/1.

\bibliography{Bibliography}
\end{document}